\documentstyle[12pt]{article}

\begin{document}
\begin{center}
{\Large {\bf Fermionic anticommutators for open superstrings
 in the presence of antisymmetric tensor field}}\\
\vspace{1cm} 

{\large Nelson R. F. Braga$^{a}$  and Cresus F. L. Godinho$^{b}$}  \\
 
\vspace{0.5cm}

{\sl
$^a$ Instituto de F\'{\i}sica, Universidade Federal
do Rio de Janeiro,\\
Caixa Postal 68528, 21945-970  Rio de Janeiro, Brazil\\[1.5ex]

$^b$ Centro Brasileiro de Pesquisas F\'{\i}sicas, Rua Dr Xavier Sigaud 150,\\
 22290-180,
Rio de Janeiro RJ Brazil }
\end{center}
\vspace{1cm}

\abstract  
We build up the anticommutator algebra for the fermionic coordinates
of open superstrings attached to  branes with antisymmetric tensor fields.
We use both Dirac quantization and the symplectic 
Faddeev Jackiw approach. In the symplectic case we find a way of generating 
the boundary conditions as zero modes of the symplectic matrix by
taking a discretized form of the action and adding terms that vanish
in the continuous limit. This way boundary conditions can be handled 
as constraints.   
  

\vspace{1cm}

\vspace{1cm}

\noindent braga@if.ufrj.br; godinho@cbpf.br

\vfill\eject

\section{Introduction}

Non commutativity of space time and its consequences for quantum field theory
have been one of the main objects of interest for theoretical particle physicists
in the last years.  
A general discussion and an important list of references can be 
found in\cite{SW}. 
An important source of non commutativity of space in string theory\cite{HV,CH1}
is the presence 
of an antisymmetric constant tensor field along the D-brane\cite{Po} world volumes
(where the string endpoints are located). 
The quantization of strings  attached to branes involves mixed (combination of Dirichlet
and Neumann) boundary conditions. This makes the quantization procedure more subtle 
since the quantum commutators/anticommutators must be consistent with these
boundary conditions. 
For the bosonic string coordinates, the non commutativity at end points has 
already received much attention. Many important aspects have been discussed 
and the commutators have been explicitely calculated (see for example 
\cite{1}-\cite{7}). 

In contrast, the complete canonical structure (anticommutators) for the 
fermionic coordinates when antisymmetric tensor fields are present 
has not yet been presented explicitely, although some important aspects 
have already been discussed\cite{1,a,b}.
As we will see here, the requirement of consistency with boundary conditions
affect the structure of the anticommutation relations at the string endpoints
even in the absence of any external field. 
We start calculating the complete Dirac
anti-brackets for the fermionic coordinates consistent with  the 
boundary conditions of strings attached to branes with anti-symmetric tensor field.
We will do this by introducing a discretization along the string spacelike coordinate. 
Such a construction make it transparent the behavior of the anticommutators 
at the end points. The string boundary conditions will lead to a
discontinuity in the anticommutator at the endpoints. Also it will emerge that,
in contrast to the standard canonical form, the anticommutator 
between the fermionic components $\,\psi^\mu_{(+)} \,,\,\psi^\mu_{-}\, $ is not vanishing 
at the endpoints  even in the absence of the antisymmetric field.

Then we consider the symplectic quantization scheme and  
develop a procedure of generating the fermionic string boundary conditions 
as constraints directly from the symplectic matrix.  
We will consider again a discretization  of the string world sheet spatial coordinate.
In a previous article \cite{7} we found the boundary conditions for the bosonic string
from the corresponding symplectic matrix  by  means of some field redefinitions.  
Here we will improve such a procedure  making use of the  fact that a 
finite number of terms that 
vanish in the continuous limit may be added to the discretized form of the action.
By choosing appropriate terms we will get the boundary conditions as zero
modes of the fermionic symplectic matrix.  Then the anticommutators will be calculated 
in the standard way.

\section{The model}
 Let us start with a superstring coupled to an antisymmetric tensor field
living on a brane. Considering just the coordinates along the brane, the action
can be represented in superspace as
  
\begin{equation}
\label{1}
S \,=\,
{ - i \over 8\pi \alpha^\prime }
\int_{\Sigma} d^2\sigma \,d^2 \theta \Big(
{\overline D} Y^\mu D Y_\mu + {\cal F}_{\mu\nu} {\overline D} Y^\mu \rho_5 D Y_\mu
 \Big)
\end{equation}

\noindent where the superfield $$ Y^\mu (\sigma^a, \theta) \,=\, X^\mu (\sigma^a) 
+ {\overline \theta}  \psi^\mu (\sigma^a) + 1/2 {\overline \theta} \theta 
B^\mu  (\sigma^a) $$ 
\noindent contains the bosonic and fermionic spacetime string coordinates.
 In components the action 
reads\footnote{Our conventions are 
\begin{equation}
\rho^1 \,=\, \pmatrix{0&i\cr
i&0\cr}\,\,\,,\,\,\,\rho^0 \,=\, 
\pmatrix{0&-i\cr i&0\cr} \,\,\,,\,\,\,\rho_5 \equiv \rho^0 \rho^1 \,\,\,\,,\,\,\,\,
\partial_\pm \,=\,\partial_0 \,\pm \, \partial_1
\end{equation}
}

\begin{eqnarray}
\label{2}
S &=&
{ 1 \over 4\pi \alpha^\prime }
\int_{\Sigma} d^2\sigma \Big(
\eta_{\mu \nu } \partial_a X^\mu \partial^a X^\nu 
\,+ \, \epsilon_{ab}{\cal F}_{ij} \partial_a X^i \partial_b X^j - B^\mu B_\mu \nonumber\\
&-& i {\overline \psi}^\mu \rho^a \partial_ a \psi_\mu 
+ i {\cal F}_{\mu\nu} {\overline \psi}^\mu \rho_b \epsilon^{ab}  \partial_ a \psi^\nu 
 \Big)\,.
\end{eqnarray}

The bosonic and fermionic sectors decouple. We will consider just the 
fermionic sector once the bosonic sector was already discussed\cite{1,2,3,4,5,6,7}.
The fermions are Majorana and can be represented as

\begin{equation}
\psi^\mu \,=\, \pmatrix{\psi^\mu_{(-)}\cr \psi^\mu_{(+)}\cr}.
\end{equation}

\noindent So that the fermionic sector reads
\begin{equation}
\label{2}
S_0 \,=\,
{-i \over 4\pi \alpha^\prime }
\int_{\Sigma} d\tau d\sigma \Big(
\psi^\mu_{(-)}  \partial_+\, \psi_{(-)\,\mu }\,+\,
\psi^\mu_{(+)}  \partial_- \, \psi_{(+)\,\mu}\,
\,-\, {\cal F}_{\mu\nu} \psi^\mu_{(-)}  \partial_+ \, \psi^\nu_{(-)}
+ {\cal F}_{\mu\nu} \psi^\mu_{(+)}  \partial_- \, \psi^\nu_{(+)} \,\Big)
\end{equation}

The minimum action principle $\delta S \,=\,0\,$ leads to a volume term
that vanishes when the equations of motion hold and also to a surface term:

\begin{equation}
\label{3}
\Big( \psi^\mu_{(-)} (\eta_{\mu\nu} - {\cal F}_{\mu\nu}) \delta \psi^\nu_{(-)}\,-\,
\psi^\mu_{(+)} ( \eta_{\mu\nu} +  {\cal F}_{\mu\nu}) \delta \psi^\nu_{(+)}
\Big)\vert_{0}^\pi\,\,=\,\,0\,\,.
\end{equation}

It is not possible to find non trivial boundary conditions involving
$\psi^\mu_{(-)}$ and $\psi^\mu_{(+)}$ that  makes this surface term vanish. 
However the solution to this problem shows up when we take into account a
result  from reference \cite{a} (see also \cite{NL}). 
There it was shown that in order to keep supersymmetry unbroken
at the string endpoints it is necessary to include a boundary term to the action.
Actually,  considering the boundary term (\ref{3}), we realize that it is 
impossible even to solve the boundary condition unless some extra term 
is added to the action. 
The interesting thing is that the same kind of term 
proposed in \cite{a} in order to restore SUSY at the end points 

\begin{equation}
S_{Bound.}\,=\,
{i \over 2\pi \alpha^\prime }
\int_{\Sigma} d\tau d\sigma \Big( {\cal F}_{\mu\nu} \psi^\mu_{(+)}
\partial_-  \,\psi^\nu_{(+)} \,\Big)
\end{equation}

\noindent will make it possible to find a solution to the boundary condition. 
Adding this term to $S_0$ the total action reads

\begin{equation}
\label{2}
S \,=\,
{-i \over 4\pi \alpha^\prime }
\int_{\Sigma} d\tau d\sigma \Big(
\psi^\mu_{(-)} {\bf E}^{\nu\mu}  \partial_+\, \psi_{(-)\,\nu }\,+\,
\psi^\mu_{(+)} {\bf E}^{\nu\mu} \partial_- \, \psi_{(+)\,\nu}\,\Big)\,\,,
\end{equation}

\noindent where ${\bf E}^{\mu\nu} \,=\, \eta^{\mu\nu} \, + {\cal F}^{\mu\nu}\,$.
The corresponding boundary term coming from $\delta S = 0 $ is now 

\begin{equation}
\label{3.1}
\Big( \psi^\mu_{(-)} \,{\bf E}_{\nu\mu}\,  \delta \psi^\nu_{(-)}\,-\,
\psi^\mu_{(+)} \,{\bf E}_{\nu\mu} \, \delta \psi^\nu_{(+)}
\Big)\vert_{0}^\pi\,\,=\,\,0\,\,.
\end{equation}

\noindent This condition is satisfied imposing the constraint that 
preserve supersymmetry\cite{1}
\begin{equation}
\label{BC}
 {\bf E}_{\nu\mu}\,  \psi^\nu_{(+)} (0,\tau)\,=\,
 {\bf E}_{\mu\nu} \, \psi^\nu_{(-)} (0,\tau)\,
\end{equation}
\begin{equation}
\label{BC2}
{\bf E}_{\nu\mu}\,  \psi^\nu_{(+)} (\pi,\tau ) \,=\,\lambda
 {\bf E}_{\mu\nu} \, \psi^\nu_{(-)} (\pi, \tau )\,\,,
\end{equation}

\noindent at the endpoints $\sigma \,=\,0$ and $\sigma = \pi\,$,
where $\lambda = \pm 1 \,$ with the plus sign corresponding to Ramond
boundary condition and the minus corresponding to the Neveu-Schwarz case.
We will only consider the endpoint $\sigma = 0$ in our calculations. The results
for  $\sigma = \pi\,$ have the same form.

Now considering the total fermionic action $ S\,$ 
we want to incorporate the boundary conditions (\ref{BC}) in a quantum 
formulation of the theory.
That means: we want to calculate anticommutators that are consistent 
with these boundary conditions.
Following the approach successfully applied to the bosonic sector 
(see for example\cite{2}-\cite{7}) we will consider  a discrete version of the string
in which we replace the continuous coordinate $\sigma$ with range  $(0,\pi )$ by 
a discrete set corresponding to intervals of length $\epsilon$.
Representing the fermionic coordinates at the endpoints of the $\,N\,$ 
intervals as:
$ \psi^\nu_{0\,(-)},\psi^\nu_{1\,(-)},\,...\,,\psi^\nu_{N\,(-)}\,;
\,\psi^\nu_{0\,(+)},\psi^\nu_{1\,(+)},\,...\,,\psi^\nu_{N\,(+)}\,$,
the discretized form of the Lagrangian  reads

\begin{eqnarray}
\label{4}
L &=&
{-i \over 4\pi \alpha^\prime }
\Large( \epsilon \psi^\mu_{0\,(-)} {\bf E}_{\nu\mu}\partial_0 \psi^\nu_{0\,(-)}
\,+\, \epsilon \psi^\mu_{1\,(-)} {\bf E}_{\nu\mu}\partial_0 \psi^\nu_{1\,(-)}
+\,...\, +
\epsilon \psi^\mu_{0\,(-)} {\bf E}_{\nu\mu}{\,\psi^\nu_{1\,(-)} - \psi^\nu_{0\,(-)}\over
\epsilon} \nonumber\\ &+& \epsilon \psi^\mu_{1\,(-)} 
{\bf E}_{\nu\mu}{\,\psi^\nu_{2\,(-)} - \psi^\nu_{1\,(-)}\over \epsilon}+ ... 
+\,\epsilon \psi^\mu_{0\,+} {\bf E}_{\nu\mu}\partial_0 \psi^\nu_{0\,+}\,\,+\,
\epsilon \psi^\mu_{1\,+} {\bf E}_{\nu\mu}\partial_0 \psi^\nu_{1\,+}
+\,...\, \nonumber\\
&-& \epsilon\, \psi^\mu_{0\,+} {\bf E}_{\nu\mu}{\,\psi^\nu_{1\,+} \,- 
\,\psi^\nu_{0\,+}\over
\epsilon} - \epsilon \,\psi^\mu_{1\,+} 
{\bf E}_{\nu\mu}{\,\psi^\nu_{2\,+} - \psi^\nu_{1\,+}\over \epsilon}+ ... \nonumber\\
\end{eqnarray}

\noindent The original theory is recovered by taking the limit 
$\,\epsilon\, \rightarrow \,0\, $.

\section{Dirac quantization}

\noindent The equal time canonical antibrackets for the original continuous fermionic 
fields are:

\begin{eqnarray}
\{ \psi^\mu_{(+)} (\sigma) \,,\,\psi^\nu_{(+)}  (\sigma^\prime) \} &=& 
\{ \psi^\mu_{(-)} (\sigma) \,,\,\psi^\nu_{(-)} (\sigma^\prime ) \}\,\,=\,\,
-  2 \pi \,i\,\alpha^{\prime}  \eta^{\mu\nu} \delta (\sigma - \sigma^\prime )
\nonumber\\
\{ \psi^\mu_{(+)}(\sigma)\,,\,\psi^\nu_{(-)}(\sigma^\prime) \} &=& 0\,\,.
\end{eqnarray}

\noindent So that the canonical antibrackets for the corresponding 
discrete fermionic variables become

\begin{eqnarray}
\{ \psi^\mu_{i\,(+)}\,,\,\psi^\nu_{j\,(+)} \} &=& 
\{ \psi^\mu_{i\,(-)}\,,\,\psi^\nu_{j\,(-)} \}\,\,=\,\,
- { 2 \pi\,i\, \alpha^{\prime} \delta_{ij} \eta^{\mu\nu} \over \epsilon}
\nonumber\\
\{ \psi^\mu_{i\,(+)}\,,\,\psi^\nu_{j\,(-)} \} &=& 0\,\,\,.
\end{eqnarray}

\noindent The discrete version of the boundary condition (\ref{BC}), that we will 
impose as a constraint in the Dirac formalism, is  

$$ \Omega_\mu \,\equiv \, {\bf E}_{\nu\mu} \psi^\nu_{0\,(+)}\,-
 {\bf E}_{\mu\nu}  \psi^\nu_{0\,(-)} \,.$$

\noindent So, the matrix of constraints is

\begin{equation}
M_{\mu\nu}\,\equiv \,\{ \Omega_\mu \,,\,\Omega_\nu \} \, =\, 
{ -4 \pi \,i\, \alpha^\prime   \over \epsilon}
\,\Big( \eta_{\mu\rho} - {\cal F}_\mu^\nu {\cal F}_{\nu\rho}\,\Big)\,\,
=\,\, { -4 \pi \,i\, \alpha^\prime   \over \epsilon}
\,\Big( \,{\bf 1}\, -\,  {\cal F}^2\,\Big)_{\mu\rho}\,\,,
\end{equation}

\noindent and the Dirac (anti-) brackets are calculated in the standard way: 

\begin{equation}
\{ A \,,\,B\,\}_D \,=\, \{ A\,,\, B \} -  \{ A\,,\,\Omega_\mu \} M^{-1}_{\mu\nu}
\{ \Omega_\nu \,,\, B\, \}\,\,.
\end{equation}

\noindent For the coordinates $\psi^\mu_{i\,\pm}\,$ with $i \ne 0$, 
corresponding to points inside the string they will be equal to the Poisson 
brackets but for the boundary coordinates we get:

\begin{eqnarray}
\label{17}
\{ \psi^\mu_{0\,(+)}\,,\,\psi^\nu_{0\,(+)} \} &=& 
\{ \psi^\mu_{0\,(-)}\,,\,\psi^\nu_{0\,(-)} \}\,\,=\,\,
- {\pi \,i\, \alpha^{\prime}   \eta^{\mu\nu} \over  \epsilon}
\\
\label{18}
\{ \psi^\mu_{0\,(+)}\,,\,\psi^\nu_{0\,(-)} \} &=& -
{\pi \,i\, \alpha^{\prime}   \over  \epsilon}
\Big( \eta^{\mu\gamma} + {\cal F}^{\mu\gamma}\,\Big)
\Big(\Big[ 1 - {\cal F}^2 \Big]^{-1}\Big)_{\gamma\rho} 
\Big( \eta^{\rho\nu} + {\cal F}^{\rho\nu} \,\Big)\,\,.
\end{eqnarray}

\noindent The anticommutators (\ref{17}) agree with the results found previously 
in ref. \cite{1}.  Now we can obtain the continuous version of our results.
Once the anticommutators of $\psi^\mu_{i\,(\pm)}\,$ for $i \ne 0$
are not changed by the Dirac quantization, inside the string 
($0\le\sigma  \le \pi$) the anticommutators keep their 
canonical form. Then, for the boundary points, we use the fact
that the mapping between continuous and discrete
expressions involve the mapping of Kronecker and Dirac deltas in the following way:
$ \delta_{ij}/ \epsilon\,\Leftrightarrow \delta (\sigma_i - \sigma_j)\,$ ( note that 
expressions (\ref{17},\ref{18}) involve a factor $\delta_{00} = 1$). 
The anticommutators of the points
inside the string and on the boundary may be acommodated in one single expression if we
introduce a parameter $\beta$ such that $\beta \,=\, 1/2 $ 
for $\sigma = \sigma^\prime = 0\,$ or $\beta = 1 $ elsewhere 
The continuous limit of the Dirac anti-brackets is then   

\begin{eqnarray}
\{ \psi^\mu_{(+)} (\sigma) \,,\,\psi^\nu_{(+)}(\sigma^\prime) \} &=& 
\{ \psi^\mu_{(-)}(\sigma) \,,\,\psi^\nu_{(-)}(\sigma^\prime) \}\,\,=\,\,
- 2\,\beta\, \pi\, i\, \alpha^{\prime}   \eta^{\mu\nu} \delta (\sigma - \sigma^\prime )
\\
\label{55}
\{ \psi^\mu_{(+)}(\sigma)\,,\,\psi^\nu_{(-)}(\sigma^\prime) \} &=& -
\pi\,i\, \alpha^{\prime} 
\Big( \eta^{\mu\gamma} + {\cal F}^{\mu\gamma}\,\Big)
\Big(\Big[ 1 - {\cal F}^2 \Big]^{-1}\Big)_{\gamma\rho} 
\Big( \eta^{\rho\nu} + {\cal F}^{\rho\nu} \,\Big)\,\delta (\sigma - \sigma^\prime )
\end{eqnarray}

\noindent for $\sigma \,=\, \sigma^\prime\,=\, 0\,$ and zero elsewhere except for
the other endpoint $\sigma \,=\, \sigma^\prime \,=\, \pi$ where the same kind of relation
holds but with a sign depending on choosing Ramond
or Neveu Schwarz boundary conditions. It is important to note that the 
anticommutator (\ref{55}) does not vanish even in the absence of the 
antisymmetric tensor field. This result is consistent with the boundary 
condition (\ref{BC}) that relates $\psi^\mu_{(+)}\,$ and $\psi^\nu_{(-)}\,$  at the 
string endpoints. 

\section{Symplectic quantization}

Let us now see how the fermionic anticommutators can be calculated using the symplectic
Faddeev Jackiw quantization\cite{FJ}. We need particularly  the analysis 
of constraints and gauge symmetries in the symplectic quantization developed in 
\cite{BW,Wo,Mo}.

We consider a Lagrangian that is first order in time derivatives 
(if the original Lagrangian is not in this form one can introduce auxiliary
fields and change it to first order). 

\begin{equation}
\label{FO}
L^0 = a_k^0 ( q ) \partial_\tau q_k - V( q ) 
\end{equation}

\noindent where $q_k$ are the generalized coordinates of the system. 
For bosonic variables the symplectic matrix is defined as

\begin{equation}
\label{f0}
f^0_{kl} \, = \, { \partial a^0_l \over \partial q_k} -
{ \partial a^0_k \over \partial q_l}\,\,.
\end{equation}

\noindent If it is non singular we define the commutators of 
the quantum theory (if there is no ordering problem for the corresponding 
quantum operators) as

\begin{equation}
\lbrack A (q) , B(q) \rbrack \,=\, { \partial A \over \partial q_k} (f^0)^{-1}_{kl} 
{ \partial B \over \partial q_l}
\end{equation}

\noindent If the matrix (\ref{f0}) is singular we find the zero modes 
that satisfy $ f^0_{kl} v^{\alpha}_l \,=\,0\,$ and the corresponding constraints:

\begin{equation}
\Omega^\alpha \,=\,v^{\alpha}_l {\partial V \over \partial q_l } \approx 0
\end{equation}

\noindent Then we introduce new variables $\lambda^{\alpha}$ and add a new term to the
kinetic part of Lagrangian  

\begin{equation}
L^1 = a_k^0 ( q ) \dot q_k + {\dot\lambda}^{\alpha} \Omega^{\alpha} - V( q ) \,\equiv \,
a_r^1 ( {\tilde q} ) {\dot {\tilde q}}_r - V( q )
\end{equation}

\noindent where we introduced the new notation for the extended variables:
 $ {\tilde q}^r \,=\,  ( q^k , \lambda^\alpha ) $. We find now the new matrix 
$ f^1_{rs}$   

\begin{equation}
f^1_{rs} \, = \, { \partial a^1_s \over \partial {\tilde q}_r} -
{ \partial a^1_r \over \partial {\tilde q}_s}\,\,.
\end{equation}

If $f^1$ is not singular  we define 
the quantum commutators as 

\begin{equation}
\lbrack A (\tilde q ) , B( \tilde q ) \rbrack \,=\, 
{ \partial A \over \partial {\tilde q}_r} 
(f^1)^{-1}_{rs} 
{ \partial B \over \partial {\tilde q}_s}\,\,.
\end{equation}

\noindent This process of incorporating the constraints in the Lagrangian 
is repeated until a non singular matrix is found. 
 
In the present case we are dealing with fermionic string coordinates.
For fermionic variables $\Psi_i$ we define 

\begin{equation}
a_{\Psi_i} \,=\, {\partial L \over \partial (\partial_\tau ) \Psi_i}\,\,,
\end{equation}

\noindent as in the bosonic case, but  the symplectic matrix takes the form 

\begin{equation}
f_{\Psi_i \,\Psi_j}\,=\, {\partial a_{\Psi_j} \over \partial \Psi_i}
\,+\, {\partial a_{\Psi_i} \over \partial \Psi_j}\,.
\end{equation}

\noindent The procedure of incorporating constraints then is the same as in 
the bosonic case.

In the previous section, the boundary conditions did not show up directly
from the Dirac procedure.
That means, the method of quantization itself did not generate the boundary conditions.
We had to impose them as additional constraints.
In the symplectic approach, in contrast, we will find the boundary conditions 
from the zero modes of the symplectic matrix. We do not get this result if we 
use directly action (\ref{4}) as our starting point.
However we note that the individual terms in this discrete form of the action 
tend to zero in the limit $\epsilon \, \rightarrow \,0 $. 
So if we remove or add a finite number of them we do not change the 
continuous limit $\epsilon \rightarrow 0 $ corresponding to the original 
action of eq. (\ref{2}). However the symplectic matrix in the discrete variables changes
and this will make it possible to find the boundary conditions as zero modes of 
the symplectic matrix.
A possible way to do this is to include in the action the extra term 

\begin{equation}
\label{extra}
{ i \over 4 \pi\alpha^\prime }
 \psi^\mu_{0\,(+)} {\bf E}_{\mu\nu} \Big(
\psi^\nu_{1\,(-)}\,-\,\psi^\nu_{0\,(-)}\,\Big)\,,  
\end{equation}

\noindent that vanishes in the limit $\epsilon \rightarrow 0$ and then 
redefine the variables as 

$$ \psi^\mu_{(\pm)\,i} \,\equiv \, {{\tilde \psi}^\mu_{(\pm ) i}
\over  \sqrt{\epsilon}}\,\,\, 
(i \neq 0\,)\,\,\,\,,\,\,\,\,\,
\, \psi^\mu_{(0)\,+} \,\equiv \, {{\tilde \psi}^\mu_{(0)\,+}\over \sqrt{\epsilon}}
 \,\,\,\,\,\,, \,\,\,\,\,\,\, 
\psi^\mu_{(0)\,(-)} \,\equiv \, {\tilde \psi}^\mu_{(0)\,(-)}$$

\noindent in the Lagrangian and we will see that this will make it possible
to  generate the appropriate boundary condition that mixes the $(+)$ and $(-)$ 
components.

\noindent The symplectic matrix takes the form

\begin{equation}
\bordermatrix{&{\tilde\psi}^\nu_{0\,(+)}&{\tilde\psi}^\nu_{0\,(-)}&
{\tilde\psi}^\nu_{1\,(+)}
&{\tilde\psi}^\nu_{1\,(-)}&{\tilde\psi}^\nu_{2\,(+)}&{\tilde\psi}^\nu_{2\,(-)}&...\cr
{\tilde\psi}^\mu_{0\,(+)}&-2\,g^{\mu\nu}&0&0&0&0&0&...\cr
{\tilde\psi}^\mu_{0\,(-)}&0&-2\, \epsilon\,g^{\mu\nu}&0&0&0&0&...\cr 
{\tilde\psi}^\mu_{1\,(+)}&0&0&-2 g^{\mu\nu}&0&0&0&...\cr
{\tilde\psi}^\mu_{1\,(-)}&0&0&0&-2 g^{\mu\nu}&0&0&...\cr
{\tilde\psi}^\mu_{2\,(+)}&0&0&0&0&-2 g^{\mu\nu}&0&...\cr
{\tilde\psi}^\mu_{2\,(-)}&0&0&0&0&0&-2 g^{\mu\nu}&...\cr
...& ...&...&...&...&...&...&...\cr}
\end{equation}

\noindent (times a factor $1/i4\pi\alpha^\prime$).
In the limit $\epsilon \rightarrow 0 $ this symplectic matrix becomes singular.
The zero mode corresponds to the vector

\begin{equation}
\pmatrix{0 \cr 1 \cr 0 \cr .\cr.\cr. }\,\,,
\end{equation}

\noindent and the corresponding constraints come from 

\begin{equation}
{ \partial V \over \partial \tilde \psi^\mu_{0\,(-)} }\,=\,0
\end{equation}

\noindent Considering  the inclusion of the extra crossed term of eq. (\ref{extra})
in the potential $V$, we find that in the  $\epsilon \rightarrow 0 $
limit the constraints, returning to the original variables, are 

\begin{equation}
\Omega^\mu\,\,=\,\,
{\bf E}^{\mu\nu} \psi^\nu_{0\,(-)} \,-\,  
{\bf E}^{Tr\,\mu\nu} \psi^\nu_{0\,(+)} \,=\,0\,\,.
\end{equation}

\noindent Then we introduce a Lagrange multiplier $\lambda^\mu$ and include
the term $ {\dot \lambda}^\mu \Omega_\mu $ in the Lagrangian.
Returning to the original fermionic variables, the symplectic matrix  becomes

\begin{equation}
\bordermatrix{&\psi^\nu_{0\,(+)}&\psi^\nu_{0\,(-)}&\psi^\nu_{1\,(+)}
&\psi^\nu_{1\,(-)}&\psi^\nu_{2\,(+)}&\psi^\nu_{2\,(-)}&...&\lambda^\nu\cr
\psi^\mu_{0\,(+)}&-2\epsilon\,g^{\mu\nu}&0&0&0&0&0&...&-{\bf E}^{\mu\nu}\cr
\psi^\mu_{0\,(-)}&0&-2\epsilon\, g^{\mu\nu}&0&0&0&0&...&{\bf E}^{\nu\mu}\cr 
\psi^\mu_{1\,(+)}&0&0&-2\epsilon g^{\mu\nu}&0&0&0&...&0\cr
\psi^\mu_{1\,(-)}&0&0&0&-2\epsilon g^{\mu\nu}&0&0&...&0\cr
\psi^\mu_{2\,(+)}&0&0&0&0&-2\epsilon g^{\mu\nu}&0&...&0\cr
\psi^\mu_{2\,(-)}&0&0&0&0&0&-2\epsilon g^{\mu\nu}&...&0\cr
...& ...&...&...&...&...&...&...&0\cr
\lambda^\mu&\,-{\bf E}^{\nu\mu}&+{\bf E}^{\mu\nu}&0&0&0&0&...&0}
\end{equation}

\noindent (again times a factor $1/i4\pi\alpha^\prime$).
Inverting this matrix we find the anticommutators

\begin{eqnarray}
\label{36}
\{ \psi^\mu_{0\,(+)}\,,\,\psi^\nu_{0\,(+)}\,\}&=&
\{ \psi^\mu_{0\,(-)}\,,\,\psi^\nu_{0\,(-)}\,\}\,\,=\,
{-\pi\,i\, \alpha^\prime \over \epsilon} g^{\mu\nu}\\
\{ \psi^\mu_{0\,(+)}\,,\,\psi^\nu_{0\,(-)}\,\}&=& -
{\pi\,i\,  \alpha^{\prime}   \over  \epsilon}
\Big( \eta^{\mu\gamma} + {\cal F}^{\mu\gamma}\,\Big)
\Big(\Big[ 1 - {\cal F}^2 \Big]^{-1}\Big)_{\gamma\rho} 
\Big( \eta^{\rho\nu} + {\cal F}^{\rho\nu} \,\Big)\,\,,
\end{eqnarray}

\noindent The anticommutators (\ref{36}) agree with the previous result from
\cite{1} and reproduce the result obtained in the Dirac quantization.

\section{Conclusion}

We have calculated the fermionic anticommutators at a string endpoint 
by  Dirac and symplectic quantization. 
In both cases, the discretization of the string spatial coordinate made it 
more easy to handle the discontinuity in the antibrackets associated to 
the effect of the boundary conditions.
In the symplectic case we found a way of getting the boundary conditions from the 
symplectic matrix by adding terms that vanish in the continuous limit.

\bigskip

\noindent Acknowledgements: The authors are partially supported by  CNPq., 
FAPERJ and FUJB (Brazilian Research Agencies).

\end{document}